\documentstyle[12pt,epsf]{article}
\def\fnote#1#2{\begingroup\def\thefootnote{#1}\footnote{#2}\addtocounter
{footnote}{-1}\endgroup}

\def\inbar{\vrule height1.5ex width.4pt depth0pt}
\def\IB{\relax{\rm I\kern-.18em B}}
\def\IC{\relax\,\hbox{$\inbar\kern-.3em{\rm C}$}}
\def\ID{\relax{\rm I\kern-.18em D}}
\def\IE{\relax{\rm I\kern-.18em E}}
\def\IF{\relax{\rm I\kern-.18em F}}
\def\IG{\relax\,\hbox{$\inbar\kern-.3em{\rm G}$}}
\def\IH{\relax{\rm I\kern-.18em H}}
\def\II{\relax{\rm I\kern-.18em I}}
\def\IK{\relax{\rm I\kern-.18em K}}
\def\IL{\relax{\rm I\kern-.18em L}}
\def\IM{\relax{\rm I\kern-.18em M}}
\def\IN{\relax{\rm I\kern-.18em N}}
\def\IO{\relax\,\hbox{$\inbar\kern-.3em{\rm O}$}}
\def\IP{\relax{\rm I\kern-.18em P}}
\def\IQ{\relax\,\hbox{$\inbar\kern-.3em{\rm Q}$}}
\def\IR{\relax{\rm I\kern-.18em R}}
\def\ZZ{\relax{\sf Z\kern-.4em Z}}
\def\fnote#1#2{\begingroup\def\thefootnote{#1}\footnote{#2}\addtocounter
{footnote}{-1}\endgroup}
\def\beq{\begin{equation}}
\def\eeq{\end{equation}}
\def\bea{\begin{eqnarray}}
\def\eea{\end{eqnarray}}
\def\lleq#1{\label{#1}\eeq}

\def\notin{\ \hbox{{$\in$}\kern-.51em\hbox{/}}}

\def\a{\alpha}      \def\g{\gamma}  
     
\def\L{\Lambda}  \def\om{\omega}   
   \def\th{\theta}
   
\def\cF{{\cal F}}   
   
 \def\cO{{\cal O}}

        \def\bq{\bar q}

\def\lra{\longrightarrow}

\def\del{\partial}

\def\hbar{\bar h}

 \def\II{{\bf II}}

\begin{document}
\baselineskip=18pt
\parskip=.15truein
\parindent=0pt

\phantom{\hfill{hep--th/0010270}}

\vskip .9truein

 \centerline{\bf SCALING BEHAVIOR OF BLACK HOLE ENTROPY}

 \vskip .5truein

\centerline{\sc Rolf Schimmrigk
     \fnote{\diamond}{Email address: rks@canes.gsw.edu, netahu@yahoo.com}}

\vskip .5truein

 \centerline{\it Georgia Southwestern State University}

 \centerline{\it 800 Wheatley Street, Americus, GA 31709}

\vskip 1truein

\centerline{\bf Abstract} \vskip .1in \noindent It is shown that
the entropy of fourdimensional black holes in string theory
compactified on weighted Calabi-Yau hypersurfaces shows scaling
behavior in a certain limit. This leads to non-monotonic functions
on the moduli space.

\renewcommand\thepage{}
\newpage

\baselineskip=20pt
\parskip=.2truein
\parindent=20pt
\pagenumbering{arabic}

\section{Introduction} One of the longstanding concerns in string
theory is the problem of the vacuum degeneracy, the hope being of
identifying a dynamical principle which predicts the unique
physical ground state of the heterotic string (or M \& F-theory
\cite{jh,hw,cvone,mv}). A first step toward such a goal would be
to find physically motivated quantities which are sensitive to the
internal geometry and which define interesting functions on the
moduli space.

Natural candidates for such functions are the Yukawa couplings and
the free energy.  Both of these quantities are functions over the
($h^{(1,1)}+h^{(2,1)})$ complex-dimensional moduli space of each
individual Calabi-Yau manifold. These multidimensional components
form the connected web of the collective moduli space \cite{cdls}
and therefore even if one could compute both quantities for a
reasonable class of spaces one would have to compare sets of
functions of different order. Future progress in technology should
make this possible. To simplify this problem one can consider the
large volume limit, neglecting the instanton contributions to
these functions, and focus on the classical bulk part. In this
limit both the Yukawa coupling and the free energy can be computed
within the class of Calabi-Yau hypersurfaces embedded in weighted
projective four-space \cite{weighted}. It turns out that both
quantities show scaling behavior with respect to a scaling
variable defined by the dimension of the space of global functions
of the hyperplane bundle on the manifold, and, as a consequence,
satisfy a scaling relation between themselves \cite{rs}. The
resulting critical exponents can be viewed as specific
characteristics of this space of Calabi-Yau manifolds. Both
functions considered in \cite{rs} turn out to be monotonic and
therefore it is of interest to search for different types of
characteristics of the moduli space.

Recently progress has been made in the understanding of the
entropy of black holes in string theory \cite{sv}. In the present
note it is pointed out that the entropy of N$=$2 black holes in
type II compactifications on Calabi-Yau threefolds CY$_3$ leads to
functions with minima on the collective moduli space.

\section{Black Holes on Calabi-Yau Threefolds}

N$=2$ Calabi-Yau black holes embedded in type IIA string theory
can also be viewed as solutions of M-theory compactified on
CY$_3\times $S$^1$. In these theories there are $(h^{(1,1)}+1)$
electric charges $(q_0,q_A)$ associated to D0-branes and the
D2-branes wrapped around the $h^{(1,1)}$ 2-cycles in the
threefold. Dual to these are the magnetic charges $(p^0, p^A)$ of
the D6-brane wrapped around the threefold and the D4-branes
wrapped around the 4-cycles. In the large volume limit the entropy
of black holes which carry only 0-brane and 4-brane charges $(q_0,
p^A)$ the entropy has been found in \cite{fks,as,fk,bcdklm,rey} to
be determined by the central charge of the theory by solving the
Ferrara--Kallosh equation \cite{fk}. The general solution of this
extremization condition for the central charge has not been found
yet but for a number of types of black holes this relation has
been solved. Useful in the present context are axion--free black
holes for which the entropy is determined by the classical Yukawa
couplings $C_{ABC} = \int_{{\rm CY}_3} \om_A\om_B\om_C,$ where
$\om_{\L} \in $H$^2($CY$_3$), and the linear form of the first
Pontrjagin class. Interesting probes of the moduli space can be
provided by black holes with nonvanishing charges $(q_0, p^A)$.
For such black holes the entropy is given by \cite{bcdklm} \beq S
\sim \frac{2\pi}{\sqrt{6}} \sqrt{q_0 C_{ABC}p^Ap^Bp^C} \eeq
 for large charges. The
microscopic derivation of this entropy as a gas of D0-branes
obtained from intersecting D4-branes has been described in refs.
\cite{bm,jm}.

As will become clear in the following section, a more interesting
function on the moduli space can be obtained by taking into
account a shift in the theta angle. It was shown in \cite{bcdklm}
that such a shift leads to an entropy of black holes with D0-brane
and D4-brane charges in the large volume limit which is of the
form \beq S \sim \frac{\pi}{6} \sqrt{\left(24q_0+c_2 J_A
p^A\right) C_{ABC}p^Ap^Bp^C}.\lleq{yukscale}
 The microscopic origin of this shift is to be found in the
additional D0-brane contribution coming from the anomaly in type
IIA string theory \cite{vw,bsv}
 as pointed out in \cite{jm}.

For fixed charges these are complicated functions on each of the
components of the collective moduli space
 which will have a rather involved, perhaps unilluminating, behavior.

\section{Scaling behavior of Black Hole Entropy}

In order to obtain simpler probes of the moduli space it is more
useful to focus on the bulk contribution to the entropy which
originates from the universal deformation defined by the
restriction of the hyperplane bundle of the ambient space.
Denoting $q_0\equiv q$ and $p^1 \equiv p$ and $C_{111}\equiv C$ we
focus on black holes with electric-magnetic charges $(p,q)$ such
that the entropy simplifies to \beq  S_{p,q} \sim 2\pi \sqrt{qCp}.
\lleq{simpleentr} Considering anti-D0-branes of charge $q_0=-q$
the anomaly shifted entropy in turn becomes \beq S^{\th}_{p,q}
\sim \frac{\pi p^2}{6}
       \sqrt{\left(-24\frac{q}{p}+c_2\cdot L \right)C}.\eeq

The Yukawa couplings of the universal K\"ahler deformation induced
by the ambient K\"ahler form is given by the degree
 the degree of a natural line bundle which on hypersurfaces embedded
 in weighted projective space is induced by the hyperplane bundle of the
 ambient space. This leads to
\beq C \equiv \int_M c_1^3(L). \eeq These couplings were computed
in \cite{rs} for the class of weighted Calabi-Yau hypersurface
threefolds constructed in \cite{weighted}. For such spaces the
 natural candidate for
a line bundle is the pullback of the weighted form of the
hyperplane bundle \beq L = \cO^{(k)}_{\IP_{(k_1,...,k_5)}} \eeq on
the weighted ambient space \cite{bk93,d93} \beq
k=lcm\{~\{gcd(k_i,k_j)|~i,j=1,..,5; i\neq j\} \cup \{k_i| k_i ~
{\rm does~not~divide}~ \sum_{i=1}^5 k_i \} \}. \eeq The pullback
$j^*(\cO^{(k)}_{\IP_{(k_1,...,k_5)}})$ of $L$ from the ambient
space to the embedded Calabi--Yau manifold $j: M \lra
\IP_{(k_1,...,k_5)}$ induces an antigeneration $j^*(c_1(L))$,
which will also be denoted by $L$. The Yukawa coupling $C(L)$ of
this antigeneration leads, in the large radius limit, to the
expression \beq C = \int_M (j^*(c_1(
\cO^{(k)}_{\IP_{(k_1,...,k_5)}}))^3
      =\left(\frac{\sum_{i=1}^5 k_i}{\prod_{i=1}^5 k_i}\right)~k^3.
\eeq

 The result shows that for
this class this Yukawa coupling defines a scaling type function on
the collective moduli space which is of the form $$C \sim
\frac{6h-16}{ 1+a h^{-\a}},$$ where $h={\rm dim~H}^0($CY$_3,L)$ is
the dimension of the space of global functions on the individual
components of the large moduli space. This dimension can be
computed for the bundles $\cO^{(k)}_{\IP_{(k_1,...,k_5)}}$ as \beq
h^0( \cO^{(k)}_{\IP_{(k_1,...,k_5)}}) = \frac{1}{k!}
\frac{\del^k}{\del t^k}
      \left(\frac{\left(1-t^{\sum_i k_i}\right)}{\prod_i \left(1-t^{k_i}\right)}
     \right)~\rule[-4mm]{.1mm}{10mm}_{~t=0}.
\eeq The constants appearing in this scaling relation are
approximately $(a,\a)=(5,0.7)$. This leads for singly charged
black holes to an entropy at zero theta angle which scales for
large couplings as \beq S_{p,q} \sim s_{p,q}\
h^{1/2},\lleq{simplescale} where $s_{p,q}$ is constant.

The entropy at nonvanishing theta angle again leads to a more
interesting scaling form by using the second scaling relation
described in \cite{rs} for the linear form defined by the second
Chern class on the second cohomology group \beq c_2\cdot L \equiv
\int_M c_1(L)\wedge c_2(M) \lleq{c2} where $c_2(M)$ is the second
Chern class of the Calabi--Yau 3--fold. This number has been of
relevance in \cite{bcov93} where it was shown that the generalized
index introduced in \cite{cfiv92} \beq \cF =\frac{1}{2} \int
\frac{d^2 \tau}{\tau_2}~
         Tr\left[(-1)^F F_L F_R q^{H_L} \bq^{H_R}\right].
\eeq describes the one--loop partition function of the twisted
N$=$2 theory coupled to gravity. Here the integral is over the
fundamental domain of the moduli space of the torus, $F_{L,R}$
denote the left and right fermion numbers and the trace is over
the Ramond sector for both the left-- and right--movers
\fnote{1}{The contribution of the ground states of the
supersymmetric Ramond sector
         to $\cF$  has to be deleted in order for the integral to converge.}.
It was shown in \cite{bcov93} that this generalized index reduces
in lowest order to \beq \cF^{\uparrow} = \frac{1}{24} \int_{M}
K\wedge c_2(M) \lleq{freetop} where $K$ is the K\"ahler form of
the manifold. Thus the numbers (\ref{c2}) define the universal
contribution $\cF^{\uparrow}_L = L\cdot c_2/24$ to the large
radius limit of this partition function.

 The resulting  scaling
relation now takes the form \beq
 c_2\cdot L \sim b\ C^{\beta},\lleq{ctwoscale}
  where $(b,\beta) = (36, 0.3)$.  Thus the entropy for large
coupling scales like \beq
 S^{\th}_{p,q}(C) \sim \frac{\pi^2 p^4}{ 36}
          \sqrt{\left(-\frac{24q}{ p}+b
          C^{\beta}\right)C}. \lleq{thetascale}
For $q/p << C$ this leads to the critical exponent $\g_C =
\frac{\beta +1}{ 2}.$

More interestingly, the shifted black hole entropy  defines a
probe which leads to nonmonotonic functions on the moduli space.
Considering for convenience the square of the entropy one finds
finds that for fixed charges $ S^2_{(1,1)}(C)$ takes its minimum
value at \beq C_{min} = \left(\frac{q}{ p}\frac{24}{ b(\beta
+1)}\right)^{\frac{1}{\beta}}. \lleq{scalemin} The behavior of the
function $ S^{\th}_{p,q}(C)$ is illustrated in Figure 1.

\vskip .3truein
\centerline{\epsfbox{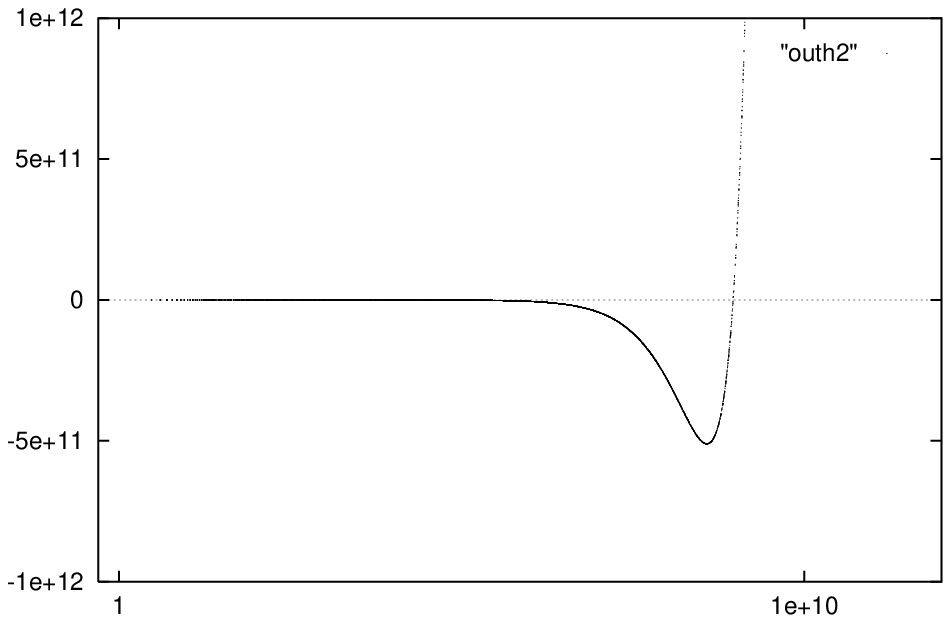}}

\noindent \centerline{ {\bf Figure 1:} {\it $(S^{\th}_{q,p})^2$ at
$q/p=10^3$ for the class of
      hypersurface threefolds \cite{weighted}.}}

Combining the two scaling relations \ref{yukscale} and
\ref{ctwoscale}  leads to the following behavior of the square of
the entropy as a function of the scaling variable $h$ \beq
(S^{\th}_{p,q})^2(h) \sim \frac{\pi^2p^4}{ 6}
      \left[-\frac{q}{ p} +\frac{b}{ 24}
               \left(\frac{6h-16}{ 1+a h^{-\a}}\right)^{\beta}
       \right]\frac{6h-16}{ 1+a h^{-\a}},\lleq{dimscale}
        with asymptotic exponent $\g_h = 1+\beta.$

In summary the entropy of black holes in string theory leads to
the first non-monotonic probes of the collective moduli spaces of
Calabi-Yau vacua.

\noindent {\bf Acknowledgement:}

Part of this work was done during my stays at the Physics
Institute of the University of Bonn, the Theory Group of the Dept.
of Physics at UT Austin, and the Dept. of Physics and Astronomy at
IU South Bend. I'm grateful to these groups for hospitality and
their members for discussions. This work was supported by the NATO
grant CRG 9710045.

\end{document}